\documentclass[11pt, draftclsnofoot, onecolumn]{IEEEtran}

\usepackage{cite,graphicx,amsmath,amssymb}
\usepackage{subfigure}
\usepackage{citesort}
\usepackage{fancyhdr}
\usepackage{mdwmath}
\usepackage{mdwtab}
\usepackage{balance}
\usepackage{xcolor}
\usepackage{bm}
\usepackage{amsthm}
\usepackage{algorithm}
\usepackage{algorithmic}
\usepackage{multirow}
\usepackage{flafter}

\usepackage{amssymb}
\usepackage{amsmath}
\usepackage{cite}
\usepackage{url}
\usepackage{xcolor}
\usepackage{cite,graphicx,amsmath,amssymb}
\usepackage{subfigure}
\usepackage{citesort}
\usepackage{fancyhdr}
\usepackage{mdwmath}
\usepackage{mdwtab}
\usepackage{caption}
\usepackage{amsthm}
\usepackage{lipsum}

\graphicspath{{./fig/}}

\begin{document}
\title{Multiple Antenna Assisted Non-Orthogonal Multiple Access}
 \markboth{\textit{A Manuscript Accepted by The IEEE   Wireless  Communications} }{}

\date{\today}

\author{


 Yuanwei~Liu,~\IEEEmembership{Member,~IEEE,}
        Hong~Xing,~\IEEEmembership{Member,~IEEE,}
        Cunhua Pan,~\IEEEmembership{Member,~IEEE,}
        Arumugam~Nallanathan,~\IEEEmembership{Fellow,~IEEE,}
        Maged~Elkashlan,~\IEEEmembership{Member,~IEEE,}
         and Lajos~Hanzo,~\IEEEmembership{Fellow,~IEEE}

\thanks{ This work was supported in part by the U.K. Engineering and Physical Sciences Research Council (EPSRC) under Grant EP/N029666/1.}
\thanks{Y. Liu, H. Xing, C. Pan, A. Nallanathan, and M. Elkashlan are with Queen Mary University of London, London,
UK (email: \{yuanwei.liu, h.xing, c.pan, a.nallanathan, maged.elkashlan\}@qmul.ac.uk).}
 \thanks{L. Hanzo is with University of Southampton, Southampton,
UK (email:lh@ecs.soton.ac.uk).}}
 \maketitle

\begin{abstract}
Non-orthogonal multiple access (NOMA) is potentially capable of circumventing the limitations of the classic orthogonal multiple access schemes, hence it has recently received significant research attention both in industry and academia.  This article is focused on exploiting multiple antenna techniques in NOMA networks, with an emphasis on investigating the rate region of multiple-input multiple-output (MIMO)-NOMA, whist reviewing two popular multiple antennas aided NOMA structures, as well as underlining resource management problems of both single-carrier and multi-carrier MIMO-NOMA networks. This article also points out several effective methods of tackling the practical implementation constraints of multiple antenna NOMA networks. Finally, some  promising open research directions are provided in context of multiple antenna aided NOMA.
\end{abstract}

\section{Introduction}
Given the popularity of bandwidth-thirsty multimedia applications, such as online gaming and virtual reality, the bandwidth demand for high-rate services has  been higher. Moreover, the proliferation of Internet of Things (IoT) devices imposes additional challenges on the next generation networks. Non-orthogonal multiple access (NOMA), is potentially capable of improving the  spectral efficiency whilst supporting the connectivity of a myriad of devices~\cite{Liu2017Pro}. Hence NOMA has been considered as a promising candidate for the fifth generation (5G) networks~\cite{Dai2015NOMA}. The key concept of NOMA relies on allowing multiple users to occupy the same resource block, whilst identifying users based on their different power levels . More particularly, NOMA applies superposition coding (SC) at the transmitters for multiplexing users within the power domain and invokes successive interference cancellation (SIC) at receivers for detection.

We focus our attention on investigating the family of multiple-antenna aided NOMA systems, since it is important to  explore the spatial degrees of freedom for improving the spectral efficiency. A remarkable advantage of a multiple-antenna aided NOMA design is that it is capable of providing array gains by invoking directional beamforming or by increasing the system's throughput by applying spatial multiplexing. Moreover, another astute application of multiple antennas in NOMA systems is based on creating unique, user-specific channels by adopting appropriate transmit precoding matrix designs. These extra manipulations are capable of eliminating the channel difference constraints of NOMA, which leads to a generalized NOMA design for satisfying the heterogeneous  quality-of-service (QoS) requirements of users. Although as mentioned above there are potential benefits, when applying multiple antennas in NOMA, numerous open research challenges arise, which motivate us to develop this paper.

The main issues in the context of multiple antenna NOMA  addressed by this article addressed are summarized as follows. 1) The rate region gains of multiple-input multiple-output (MIMO)-NOMA compared to MIMO-OMA are presented from a theoretical perspective. 2) A pair of representative multiple-antenna aided designs are illustrated. Based on those designs, the resource allocation problems of multiple-antenna aided NOMA are discussed in terms of both single-carrier and multi-carrier forms. 3) The associated practical implementation issues were identified and a range of promising solutions were discussed.


\section{Multiple-Antenna NOMA: Information Theoretic Perspective}
In this paper, we confine our discussions of the multiple-antenna NOMA downlink to specific types of multi-user broadcast channels (BCs). In this section, we discuss the fundamental limits of the multi-user BC unveiling that NOMA is capable of improving the downlink spectral efficiency, because information theoretically it is optimal in terms of its achievable rate region in several important special cases. 


\subsection{MISO-NOMA}

It is widely recognized that the capacity region of a {\em degraded} BC is achievable by using superposition coding at Transmitters (Txs) and SIC at a receivers (Rxs). Specifically, assuming a two-user single-input single-output (SISO) case with the users ordered naturally by their channel gains, e.g., \(\vert h_n\vert^2>\vert h_m\vert^2\), the conditions of \(R_{n\rightarrow m}>R_{m\rightarrow m}\)\footnote{\(R_{i\rightarrow j}\) denotes the rate at which user \(i\) decodes user \(j\)'s message throughout the paper.} that guarantees successful SIC is {\em automatically} satisfied.

On the other hand, assuming a base station equipped with more than one transmit antenna serving two users each associated with \(r_i\) (\(i=m,n\)) receive antennas, even when we have \(r_m=r_n=1\),  there is no known results on its capacity region of this general Gaussian multiple-input single-output (MISO) BC. Hence it is  only that the dirty paper coding (DPC) rate region is recognized as being achievable, which is in general larger than NOMA's rate region assuming the same fixed DPC encoding and SIC decoding order. The reason that accounts for the reduction of the NOMA rate compared to DPC is that, for two-user MISO downlink transmission assuming a fixed order of \(m\rightarrow n\), DPC ensures that the achievable rate of User $n$ is ${R_n} = {\log _2}(1 + |{\bf{h}}_n^{\bf{H}}{{\bf{w}}_n}{|^2})$, where ${{\bf{h}}_n}$ and ${{\bf{w}}_n}$ are the channel vectors and beamformer vectors of User $n$, respectively. This is because the interference caused by User \(m\) has been assumed to be non-causally known and thus has been pre-cancelled by the Tx. By contrast, NOMA whose rate region is achieved by SIC entails the extra constraint of \(R_{n\rightarrow m}>R_{m\rightarrow m}\) so that once the message of User \(m\) is successfully decoded, it can also be successfully remodulated and cancelled by User \(n\). As a result, the rate region achieved by NOMA is contained with that achieved by DPC. Analogous to the SISO case, one may assume another natural ordering of the users, according to \(\|\bf h_n\|>\|\bf h_m\|\), which however does not necessarily yield \(R_{n\rightarrow m}>R_{m\rightarrow m}\), as it can be readily verified by simple calculation.

It is also intriguing to find in the literature that in some {\em special cases}, MISO NOMA is capable of achieving the same performance as DPC. For instance, the sufficient and necessary conditions for a {\em quasi-degradation} was recently developed in \cite{Z.chen16} in order to bridge the gap between NOMA and DPC.

\subsection{MIMO-NOMA}
Similar to the MISO case, the capacity region of a general MIMO downlink transmission is still unknown, while the DPC rate region coincides with the capacity region of the MIMO BC in several special cases, such as that of the {\em aligned} and {\em degraded} MIMO BC (ADBC), and that of the {\em aligned MIMO BC} (not necessarily degraded) \cite{H.Weingarten06}. More particularly, it was shown in \cite{H.Weingarten06} that  the NOMA rate region under a covariance matrix input constraint of \(\bf S\) (\(\bf S\succeq\bf 0\)) is readily achievable for the ADBC. Furthermore, it was also shown that in this case we have capacity region \(=\) DPC rate region \(=\) NOMA rate region. The relations among the rate regions achieved by the different transmission schemes considered are summarized in Table~\ref{table:relations among different regions}.

\begin{figure}[!htp]
\begin{center}
\includegraphics[width=12cm]{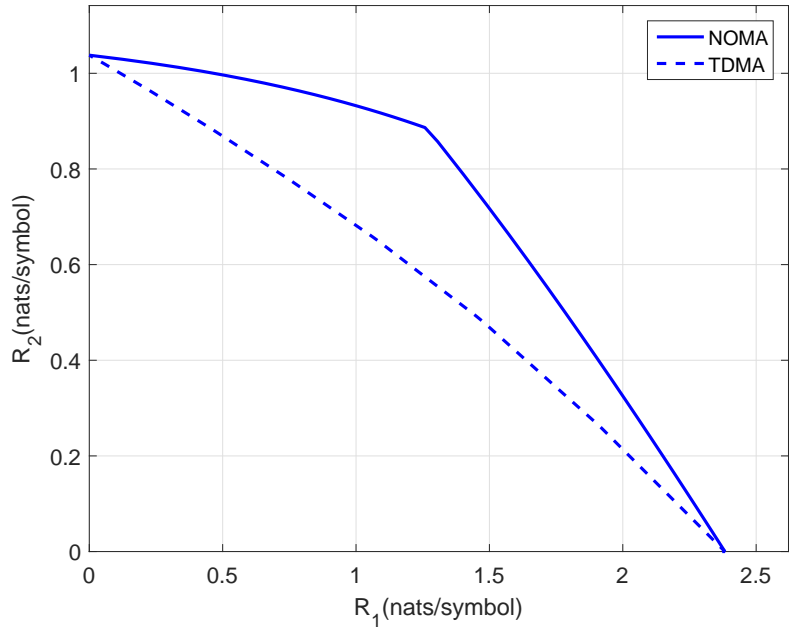}
 \end{center}
\caption{The boundaries of the two-user ADBC rate regions of both NOMA and of TDMA, in conjunction with \({\bf N_1}=[.5,.18;.18,.7]\), \({\bf N_2}=[.7, .08;.08, 10.7]\) and \({\bf S}=[1, .6;.6,2]\). Here, $\bf N_i$, $i=1,2$, denotes the additive noise covariance matrix at the $i$th Rx, and $\bf S$ is the given matrix covariance matrix.}\label{fig:ADBC rate region}
\end{figure}

We continue by providing a numerical example for the ADBC downlink. The MIMO BC is said to be aligned if the number of transmitt antennas is equal to the number of receive antennas at each of the Rxs, i.e.,  we have \(t=r_m=r_n\), and the channel gain matrices are all identity matrices; By contrast, the MIMO BC is said to be degraded if the covariance matrices of the additive Gaussian noise at each of the Rxs are ordered as \(\bf 0\prec\bf N_n\preceq\bf N_m\)\footnote{\(\bf A\preceq\bf B\) denotes $({\bf{A}} - {\bf{B}})$ is a negative semi-definite matrix.}. We compare in Fig.~\ref{fig:ADBC rate region} the downlink NOMA rate region, i.e., the capacity region, of a two-user \(2\times 2\) ADBC against that achieved by an OMA scheme, namely the classic time-division multiple access (TDMA). 



\section{Multiple-Antenna NOMA: Beamformer Based Structure}

While previous section has laid the foundation for multi-antenna NOMA from a theoretical perspective, in  the following sections we will discuss the promising multiple-antenna aided NOMA solutions.   Broadly speaking, current applications can be primarily classified into a pair of categories, namely, beamformer based structure and cluster based structure. The key difference between these two structures is that whether one beamformer serves multiple users or one user. In this section, we will take our focus on  investigating each beamformer design for each single user, both by invoking the centralized and coordinated beamforming approaches.

\subsection{Centralized Beamforming}

\begin{figure*}[!ht]
\centering
\subfigure[Illustration of centralized beamforming of multiple-antenna aided NOMA.]{\label{centralized BF}
\includegraphics[width=12cm]{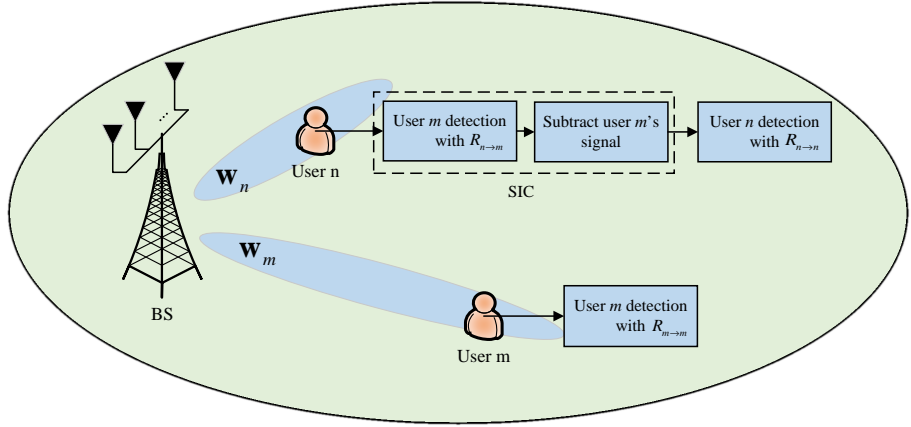}}
\subfigure[Illustration of coordinated beamforming of multiple-antenna aided NOMA.]{\label{Coordinated_BF}
\includegraphics[width=12cm]{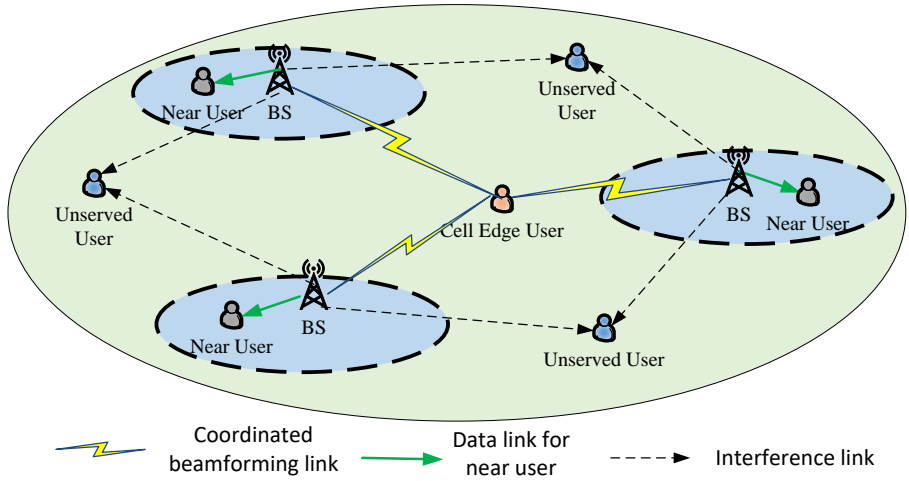}}
\caption{Beamformer based structure of multiple-antenna aided NOMA}\label{BF_NOMA}
\end{figure*}

The centralized beamforming terminology is used in this paper to indicate that the transmit precoding (TPC) is carried out by a single base station (BS) and there is no coordination among at the BS. In the centralized beamforming the BSs are equipped with $M$ antennas. As shown in Fig. \ref{centralized BF}, let us consider a simple two-user dnowlink MISO scenario as our example, where the BS transmits a superposition  of individual messages of User $m$ and User $n$ with the aid of two beamformers ${{{\bf{w}}_m}}$ and ${{{\bf{w}}_n}}$ specifically constructed for each user. We denote the channel vectors of User $m$ and User $n$ by ${{\bf{h}}_m}$ and ${{\bf{h}}_n}$, respectively. User $n$ decodes the message of User $m$ at the rate of ${R_{n \to m}} = {\log _2}( {1 + \frac{{{{\left| {{\bf{h}}_n^H{{\bf{w}}_m}} \right|}^{^2}}}}{{{{\left| {{\bf{h}}_n^H{{\bf{w}}_n}} \right|}^{^2}} + {\sigma ^2}}}} )$, where ${\sigma ^2}$ is the noise variance. If this operation is successful, User $n$ invokes the classic SIC and then decodes its own message at the rate of ${R_{n \to n}} = {\log _2}( {1 + \frac{{{{\left| {{\bf{h}}_n^H{{\bf{w}}_n}} \right|}^{^2}}}}{{{\sigma ^2}}}} )$. As for User $m$, it will directly decode its own message by treating the message of User $n$ as interference, which is given by ${R_{m \to m}} = {\log _2}( {1 + \frac{{{{\left| {{\bf{h}}_m^H{{\bf{w}}_m}} \right|}^{^2}}}}{{{{\left| {{\bf{h}}_m^H{{\bf{w}}_n}} \right|}^{^2}} + {\sigma ^2}}}})$.

As mentioned in Section II, it is worth pointing out that  successful SIC can only be guaranteed, if the condition of ${R_{n \to m}} > {R_{m \to m}}$ is satisfied. Nevertheless, this constraint will make the resultant optimization problem very challenging to solve, since some existing research contributions of MISO NOMA rely on alternative approaches to circumvent this issue, such as assuming a significantly different path-loss for the users or assigning a predefined decoding order for each user. Furthermore, finding the optimal ordering for MISO NOMA is still an open problem, hence further research efforts are expected to find the optimal performance bound.


\subsection{Coordinated Beamforming}

In addition to the centralized beamforming approach, coordinated beamforming is another effective technique of invoking multiple-antenna technologies. The concept of coordinated beamforming relies on the cooperation of a number of single antenna devices for the sake of forming virtual antenna arrays, which however relies on setting aside much of the achievable capacity for inter-node information exchange. Fig.~\ref{Coordinated_BF} illustrates a possible implementation of coordinated beamforming in NOMA. More particularly, several BSs are engaged in coordinated beamforming to serve a  cell edge user, where each BS serves a single user within its own cell by applying SIC for cancelling the intra-cell interference received from the cell-edge user. By doing so, the performance of the cell edge user is enhanced, which results in better fairness for the entire network.


The initial idea of coordinated beamforming of NOMA, which is similar to the coordinated multi-point (CoMP) transmission scheme, was proposed in \cite{Jinho2014Comp}, where two coordinated BSs invoke an Alamouti code based coordinated SC to simultaneously serve a pair of users in each others' vicinity as well as a cell edge user. As an evolution of the single antenna based systems of \cite{Jinho2014Comp}, the authors also considered multiple-antenna assisted BSs and users in \cite{Shin2017coordinated}, which was essentially a coordinated MIMO-NOMA system setup. In particular, a joint centralized and coordinated beamforming design was developed for suppressing the inter-cell interference as well as for enhancing the throughput of the cell edge users.
\section{Multiple Antenna NOMA: Cluster Based Structure}
Another popular multiple-antenna NOMA design relies on partitioning the users into several different clusters, where the users in a specific cluster share the same beamformers. Then with applying appropriate TPC and detector designs, the inter-cluster interference can be suppressed. In this section, we introduce two typical cluster based designs, depending on whether the inter-cluster interference can be completely cancelled.

\subsection{Inter-Cluster Interference Free Design}
\begin{figure*}[!ht]
\centering
\subfigure[Illustration of cluster based structure for multiple-antenna NOMA.]{\label{cluster_based_design}
\includegraphics[width=12cm]{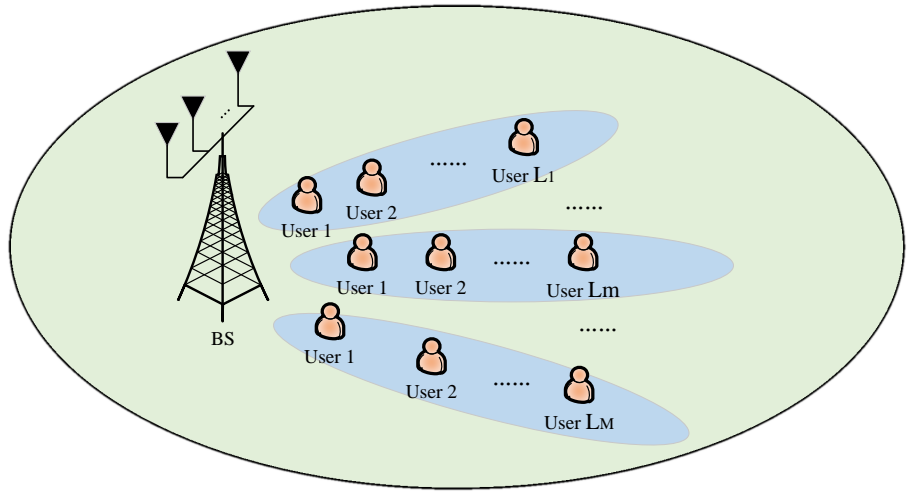}}
\subfigure[Resource allocation for guaranteeing the fairness of cluster based MIMO-NOMA networks~\cite{Yuanwei2016NOMA}. ``OP" refers to ``optimal power allocation", which is obtained by bi-section search. ``RP" refers to ``random power allocation".]{\label{resource allocation}
\includegraphics[width=12cm]{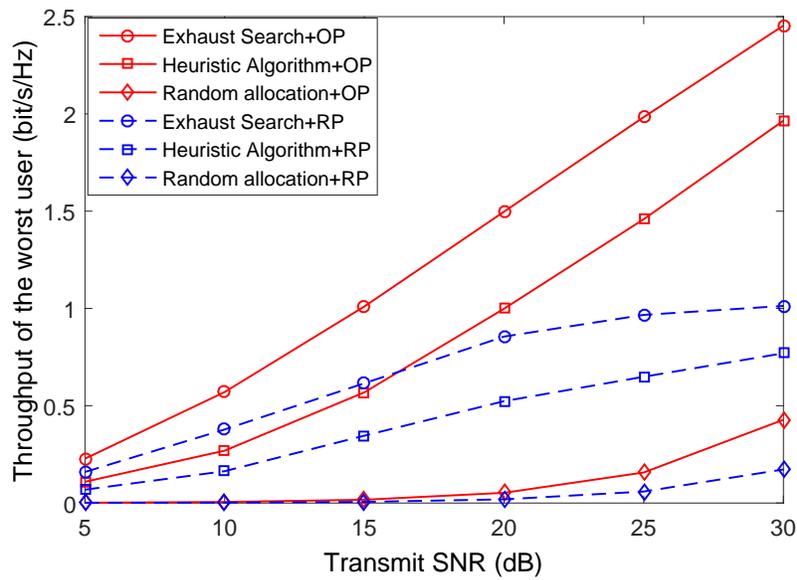}}
\caption{Cluster based structure of multiple-antenna aided NOMA}\label{cluster_based}
\end{figure*}

In an effort to tackle the channel ordering, an appealling low-complexity technique is that of decomposing the  MIMO-NOMA channels to multiple SISO-NOMA channels \cite{ding2015mimo,Zhiguo2015general}. As shown in Fig. \ref{cluster_based_design}, a BS equipped with $M$ antennas communicates with $K = \sum\nolimits_{m = 1}^M {{L_m}}$ users, who are randomly partitioned into $M$ clusters and equipped with $N$ antennas each.  The spirit of this decomposition based design is to adopt zero-forcing detection for each user, which results in a low complexity SISO-NOMA model. Hence the conventional NOMA technique can be applied.  A range of TPC designs can also be correspondingly invoked at the BS \cite{Zhiguo2015general}. 


A remarkable advantage of this decomposition aided design is that by transforming to SISO-NOMA, the sophisticated channel ordering may be circumvented, which reduces the system complexity. Moreover, this design is capable of completely canceling the inter-cluster interference.  Nevertheless, the zero-forcing detection adopted relies on a specific relationship between the number of transmitter antennas and receiver antennas. Specifically, either the condition of  $N \ge M$ of \cite{ding2015mimo} or that of $N \ge M/2$ of \cite{Zhiguo2015general} has to be satisfied. It is worth pointing out that fairness is of great significance in the cluster based NOMA structure.  Fig.~\ref{resource allocation}  characterized that the resource allocation conceived for maintaining  max-min fairness in clustered MIMO-NOMA scenarios, applying a heuristic algorithm for user scheduling and bi-section-based search for optimal power allocation. Further comparisons relying on other optimization techniques will be detailed in Table~II of Section V.

\subsection{Inter-Cluster Interference Tolerant Design}
In contrast to the inter-cluster interference free design mentioned in last subsection, we introduce another cluster based MIMO-NOMA design which was proposed in~\cite{Ali2017MIMONOMA}. This design allows the existence of inter-cluster interference and applies the so-called decoding scaling weight for increasing the strength of the desired signals. We consider Fig. \ref{cluster_based_design} as our example, and in contrast to the random clustering in~\cite{ding2015mimo,Zhiguo2015general}, the user clustering in~\cite{Ali2017MIMONOMA} follows specific techniques for making the channel gains of users more distinctive. The users within a specific cluster are sorted according to their equivalent normalized channel gain. As a result, the user experiencing the highest channel gain is the cluster-head, is capable of completely canceling all intra-cluster interference by invoking SIC. The key idea behind this detection approach is that the BS has to send the decoding scaling weight to the users prior to the data transmission process. The inter-cluster interference can be efficiently suppressed by exploiting the fact that all cluster users except for the cluster-head will estimate their own equivalent cluster channels.

The key advantage of this inter-cluster interference tolerant design is that it does not imposed any constraints at the BS and users conceiving their number of antennas. Hence such designs can be directly extended to the scenarios, where the BS is equipped with a large antenna array, as in massive-MIMO-NOMA  or NOMA millimeter wave communication scenarios \cite{Bichai2017JSAC}. Actually, a large antenna array will facilitate highly directional transmission, which in turn results in correlations among the users. This channel characteristic is ideal for the application of NOMA principles in the context of large antenna array systems.


%
%
%
%
%
%
%
%

%

\section{Resource Allocation for Multiple-Antenna NOMA}
Due to the complex nature of interference in multiple-antenna aided NOMA networks, especially in the cluster based design, one of the main challenges is to enhance both the spectral and energy efficiency by exploiting the sophisticated reuse of resources, with the aid of appropriately design  beamformers, by the opportunistic scheduling of users into different clusters by applying efficient algorithms and by intelligent power allocation. In this section, we will first introduce the resource allocation problems of multiple-antenna aided NOMA scenarios from the perspective of both single-carrier and multi-carrier solutions, and then summarize some potential mathematical modeling technique and tools conceived for tackling these problems.
\subsection{Single-Carrier Resource Allocation}
We commence our discussion on resource allocation in terms of MIMO-NOMA in a single-resource block, relying on a single carrier and the same time slot or spreading code. We still use the cluster based MIMO-NOMA structure of Fig. \ref{cluster_based_design}, as an example, where several resource allocation problems have to be carefully tackled: i) the number of clusters; ii) the number of NOMA users to allocate to each cluster; iii) which users should be assigned to which clusters; and iv) power sharing among the different clusters as well as among the users within the same cluster. Several MIMO-NOMA resource allocation contributions have considered a simplified model with a particular focus on tackling the third and fourth issues, such as fixing both the number as well as the size of clusters \cite{Ali2017MIMONOMA,Yuanwei2016NOMA}. More specifically, dynamic user scheduling and power allocation problems were optimized in \cite{Ali2017MIMONOMA} and \cite{Yuanwei2016NOMA},  aiming for maximizing the system's throughput and addressing the max-min fairness of clustered MIMO-NOMA scenarios, respectively, when utilizing particular TPC and decoder. Moreover, optimally solving all the aforementioned problems is a rather challenging problem, which constitutes a promising research direction.
\subsection{Multi-Carrier Resource Allocation}
Multi-carrier MIMO-NOMA constitutes a natural extension of single-carrier MIMO-NOMA, which relies on the multiplexing also in the frequency domain in addition to the power and spatial domains.  Multi-carrier resource allocation for MIMO-NOMA can be viewed as a hybrid NOMA resource allocation problem, which aims for simultaneously managing multi-dimensional resources. As the number of MIMO-NOMA users to be served in a single resource block increases, the intra/inter beams interference suppression techniques have to become more sophisticated, which may also require a large number of antennas at the BS. Driven by this, we can first partition the users into different sub-carrier bands, which are orthogonal to each other. In the same subband, MIMO-NOMA design are invoked. By adopting such designs, the system's implementational complexity can be significantly reduced.

Some initial research results are already available on  multi-carrier SISO-NOMA \cite{Boya2016NOMA,Sun2017TWC}, but multi-carrier MIMO-NOMA designs are still in their infancy. Note that the resource allocation of multi-carrier MIMO-NOMA requires more sophisticated design, which includes two rounds of user scheduling as well power allocation for both the sub-carriers and MIMO clusters.  It is worth pointing out that other practical forms of multi-carrier NOMA, such as sparse code multiple access (SCMA) and pattern division multiple access (PDMA) may also be combined with MIMO techniques, in the context of generalized multi-carrier MIMO NOMA resource allocation designs.

\subsection{Approaches for Resource Allocation}

Given the distinct characteristics of MIMO-NOMA channels and the interference constraints, the optimization problems formulated for resource allocation usually constitute a mixed-integer non-convex problem. There exist two popular methods for tackling this kind of problem. The \emph{first} one is based on the joint optimization of both user scheduling and of power allocation \cite{Sun2017TWC}. Representative  approaches invoked for solving the joint optimization problems can be monotonic optimization and Branch-and-Bound, which may result in globally optimal solutions.  The \emph{second} one is to decouple the user scheduling and power allocation into two sub-problems to be optimized \cite{Ali2017MIMONOMA,Yuanwei2016NOMA,Boya2016NOMA}. More particularly, matching theory can be an effective technique of moderate complexity for scheduling users \cite{Boya2016NOMA}. Regarding power allocation, geometric programming and non-cooperation games are promising tools for allocating the power in an intelligent way. Table II summarizes some representative optimization strategies, which can be potentially applied for tackling the resource allocation problem in MIMO-NOMA scenarios.

\section{Tackling the Practical Implementation Constraints of Multiple-Antenna NOMA}
Although the application  of multiple-antenna techniques in NOMA potentially enhances the performance of networks upon scaling up the number of antennas, it also imposes several implementational constraints in practical scenarios, such as a potentially overhead, sophisticated TPC and detector design, energy and security related issues, etc.  Motivated by this, in this section, we will provide several effective solutions for tackling these implementatonal constraints.

\subsection{Reducing Complexity  with Imperfect CSI}
The sophisticated TPC and detector designs of MIMO-NOMA may impose a high feedback overhead complied with tight specifications, which raise high requirements for channel estimations. Although many existing research contributions are based on idealized simplifying assumption of having perfect channel state information (CSI), in practical systems, this cannot be achieved. Hence low-complexity imperfect CSI based designs are desired for MIMO-NOMA networks. There exists two popular approaches: i) The first approach is to rely on partial CSI, such as the path loss, which does not fluctuate rapidly. ii) The second approach is to use limited feedback for reducing the system's overhead, which is particularly significant for MIMO-NOMA networks. By doing so, each user feeds back a limited number of bits to the BS, which may be the TPC codebook index.

\subsection{Tackling Energy Issues with Multi-Antenna Aided Wireless Power Transfer}
Given the fact that NOMA is capable of supporting massive conductivity, is eminently suitable for the IoT. However, the energy is severely limited in IoT scenarios, especially in wireless sensor networks where the user equipments (UEs) are usually energy constrained. This motivates the application of a new member of the energy harvesting family---simultaneously wireless information and power transfer (SWIPT), which was initially proposed for cooperative NOMA scenarios in \cite{yuanwei_JSAC_2015}. Multi-antenna techniques are capable of supporting the application of SWIPT in NOMA networks, which opens up several exciting opportunities. However, these applications also pose new challenges in terms of jointly designing the energy and information beams, which requires further research efforts in this area.
\subsection{Low-Complexity Design with Antenna Selection}
The classic antenna selection (AS) technique can be invoked as an effective solution for MIMO-NOMA networks due to the fact that it brings about two distinct benefits. The first one is that AS reduces the hardware costs imposed by the expensive radio frequency (RF) chains without losing the spatial diversity, which is the advantage of AS for conventional MIMO. The second one, which is also quite significant is that AS transforms MIMO-NOMA to SISO-NOMA in a straightforward manner, hence the sophisticated channel ordering operation of  MIMO-NOMA can be avoided, leading to an appealing performance-complexity tradeoff. It is worth pointing out that the design of effective algorithms for selecting antennas both at transmitters and receivers to strike an attractive performance-complexity tradeoff is a promising research area.

\subsection{Security Provision Relying on Multiple Antennas}
\begin{figure}[!ht]
    \begin{center}
        \includegraphics[width= 4.5in, height=2.8in]{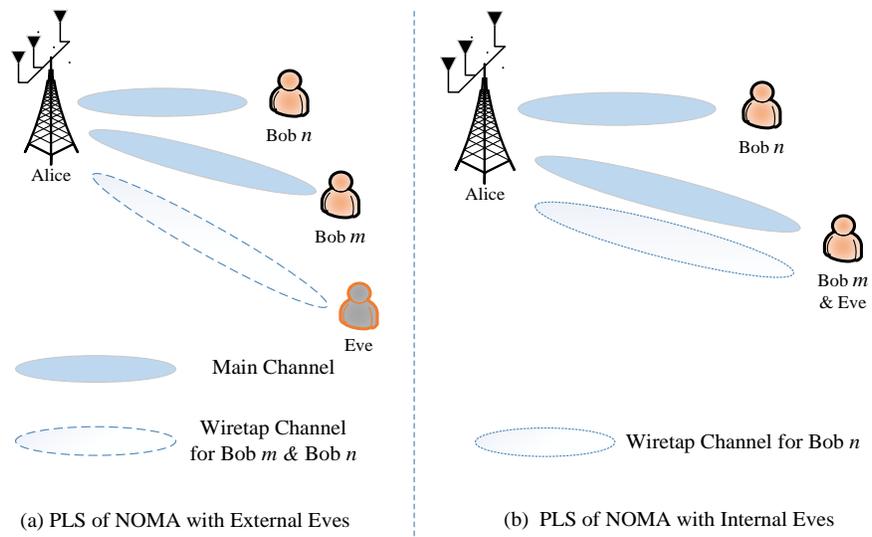}
     \caption{Illustration of PLS of Multiple Antenna NOMA.}
            \label{PLS_NOMA}
    \end{center}
\end{figure}

Invoking physical layer security (PLS) is beneficial  in NOMA networks in order to counteract the broadcast nature of wireless transmissions as well as the security threat of SIC. When dealing with security issues, exploiting multiple-antenna aided NOMA is capable of enhancing the PLS by increasing the desired destination user's capacity. Alternatively, artificial noise (AN) may be invoked on the  eavesdroppers (Eves) without degrading  the reception of the desired user (Bob). This AN aided approach was applied in a MISO-NOMA scenario, where eavesdroppers are external \cite{Yuanwei2017TWC}, as shown in Fig. \ref{PLS_NOMA}(a). By contrast, for the scenarios when the internal NOMA users are the Eves, as seen in Fig. \ref{PLS_NOMA}(b), Bob $m$ having a poor channel tries to detect the message of Bob $n$ having a good channel, multiple antennas can be used to artificially create the required effective channel differences between two users, which is beneficial for preventing eavesdropping. It is worth to point out that how to prevent Bob $n$ from detecting the signal of Bob $m$ is still an open problem.

%
%
%

\section{Conclusions and Promising Research Directions}

In this article, the application of multiple-antenna techniques to NOMA has been exploited. The capacity gain of MIMO-NOMA over MIMO-OMA has been first demonstrated from an information theoretic perspective. Then two dominant multiple-antenna NOMA structures, namely the beamformer based and the cluster based
designs have been highlighted. The resource allocation problems of MIMO-NOMA networks have also been identified, followed by discussing several implementation issues as well as the corresponding potential solutions conceived for multiple-antenna aided NOMA. There are still numerous  open research problems in this area, which are listed as follows:
\begin{itemize}
  \item \textbf{Spatial Effect Investigation}:  Stochastic geometry has been recognized as a powerful mathematical tool used for capturing the topological randomness of large-scale networks. Some initial stochastic geometry based investigations has been conducted in context of NOMA relying on the Binomial point process (BPP) and Poisson point process (PPP) \cite{yuanwei_JSAC_2015,Zhiguo2015general,Yuanwei2017TWC}, leading to the more practical but challenging  Poisson cluster process (PCP). By utilizing multiple-antenna arrays at the BS, more effective directional beamforming design can be adopted, while considering the spatial position of BSs and NOMA users on the attractive system performance, which will be a promising research direction.
  \item \textbf{Optimal Decoding Order Design}: The SIC decoding order has a significant impact on the performance of NOMA networks. Compared to the SISO-NOMA systems, the optimal ordering design problem of MIMO-NOMA is more challenging, since it depends on the TPC and on the detector design. Most of the existing research contributions on MIMO-NOMA were based on a particular ordering, which does not result in approaching the optimal performance bound. Such optimal designs still constitute an open area.
  \item \textbf{Modulation Design for MIMO-NOMA}: Given the maturity of orthogonal frequency division multiplexing (OFDM), the MIMO-NOMA designs are expected to be incorporated into OFDM.  Various other novel modulation schemes have also been proposed for 5G networks, which have to be investigated.
    \item \textbf{Emerging MIMO-NOMA for 5G}: Massive MIMO and millimeter wave, as two important technologies advocated for the forthcoming 5G networks, are capable of enhancing the attainable system performance, as a benefit of their  large antenna array gain and large bandwidth, respectively. NOMA is expected to co-exist with these two technologies for further improving the spectral efficiency as well as for supporting massive connectivity. However, the distinct characteristics of a large number antennas at the BS inevitably necessitates the redesign of TPC and detection techniques, which require more research contributions in this field.
\end{itemize}

\textbf{Yuanwei Liu} is a Lecturer (Assistant Professor) with Queen Mary University of London, where he received his Ph.D. degree in 2016. His research interests include 5G wireless networks, Internet of Things, and stochastic geometry. He received the Exemplary Reviewer Certificate of the IEEE Wireless Communication Letters and the IEEE Transactions on Communication. He serves as an Editor of the IEEE Communications Letters and the IEEE Access.

\textbf{Hong Xing}  received the B.Eng. degree in electronic sciences and technologies from Zhejiang University, China, in 2011, and the Ph.D. degree in wireless communications from King's College London, U.K., in 2015.  She was a Research Associate with the Department of Informatics, King's College London, U.K., from 2016 to 2017. She is currently a Research Associate with the School of EECS, Queen Mary University of London, U.K.

\textbf{Cunhua Pan} received his B.S. and Ph.D. degrees in Southeast University, Nanjing, China, in 2010 and 2015, respectively. From 2015 to 2016, he worked as a research associate in University of Kent, UK. He is currently a research fellow with Queen Mary University of London, UK. His research interests include ultra-dense
C-RAN, UAV, IoT, NOMA, and mobile edge computing. He serves as the Student Travel Grant Chair for ICC 2019.

\textbf{Arumugam Nallanathan} is Professor of Wireless Communications at Queen Mary University of London since September 2017.
He was with King¡¯s College London from 2007 to 2017, and National University of Singapore from 2000 to 2007. His research
interests include 5G Wireless Networks, Internet of Things and Molecular Communications. He received the Best Paper Award
at IEEE ICC 2016. He is an IEEE Distinguished Lecturer. He is a Web of Science Highly Cited Researcher in 2016.

\textbf{Maged Elkashlan} received the PhD degree in Electrical Engineering from the University of British Columbia in 2006. From 2007 to 2011, he was with the Commonwealth Scientific and Industrial Research Organization (CSIRO) Australia. In 2011, he joined the School of Electronic Engineering and Computer Science at Queen Mary University of London. Dr. Elkashlan serves as Editor of IEEE Transactions on Wireless Communications, IEEE Transactions on Vehicular Technology, and IEEE Transactions on Molecular, Biological and Multi-scale Communications.

\textbf{Lajos Hanzo} (http://www-mobile.ecs.soton.ac.uk) FREng, FIEEE, FIET, Fellow of EURASIP, DSc received his degree in electronics in
1976 and his doctorate in 1983.  He holds an honorary doctorate by the Technical University of Budapest (2009) and by the University of Edinburgh (2015).  He is a member of the Hungarian Academy of Sciences and a former Editor-in-Chief of the IEEE Press. He is a Governor of both IEEE ComSoc and of VTS.
He has published 1700+ contributions at IEEE Xplore and 18 Wiley-IEEE Press books. He has successfully supeRvised 112 PhD students.

\begin{table}[htp]
\centering
\resizebox{\textwidth}{!}{%
\begin{tabular}{llll}
\hline
\multicolumn{1}{l}{} & \multicolumn{1}{c}{Capacity region} & \multicolumn{1}{c}{DPC rate region} & \multicolumn{1}{c}{NOMA rate region} \\ \hline
SISO & \multicolumn{1}{c}{=} & \multicolumn{1}{c}{=} & \multicolumn{1}{c}{=} \\ \hline
MISO/ & \multirow{3}{*}{\begin{tabular}[l]{@{}l@{}}\(\begin{cases} =\mbox{DPC},\ \mbox{\em special cases}\\\mbox{\bf unknown},\, \ge\mbox{DPC \cite{H.Weingarten06}}.\ \mbox{in general}\end{cases}\)\end{tabular}} & \multirow{3}{*}{\begin{tabular}[l]{@{}l@{}}\(\begin{cases}
=\mbox{NOMA},\ \mbox{\em special cases, \em e.g., \em (quasi-) degraded \em\cite{Z.chen16}} \\
\ge\mbox{NOMA}.\ \mbox{in general, given the same encoding/decoding order}\end{cases}\)\end{tabular}} & \multirow{3}{*}{\begin{tabular}[l]{@{}l@{}}Assuming fixed decoding order \(m\rightarrow n\):\\successful SIC $\Leftrightarrow R_{n\rightarrow m}>R_{m\rightarrow m}$ \end{tabular}} \\
MIMO &  &  & \vspace{.2in}\\ \hline
\end{tabular}%
}\vspace{.5em}\caption{Relationships among the rate regions achieved by NOMA and others.}\label{table:relations among different regions}
\end{table}


\begin{table}[!ht]\scriptsize 
\begin{center}
{\tabcolsep12pt\begin{tabular}{|l|l|l|l|l|}\hline   
  \textbf{Categories} & \textbf{Optimization Variables}  &\textbf{Optimization Approaches} & \textbf{Characteristics} & \textbf{Ref.} \\
     \hline
\multirow{2}{*}{Jointly}    & \multirow{1}{*}{User Scheduling \&}   & $\bullet$ Monotonic Optimization & Optimality achievable  & \cite{Sun2017TWC} \\
\cline{3-5}
                     & Power Allocation & $\bullet$ Branch-and-Bound  & Optimality achievable & ---  \\
\hline
\multirow{4}{*}{Decoupled}  & \multirow{2}{*}{User Scheduling} & $\bullet$ Matching Theory& Moderate complexity & \cite{Boya2016NOMA} \\
\cline{3-5}
               &  & $\bullet$ Heuristic Algorithms & High flexibility  &  \cite{Ali2017MIMONOMA,Yuanwei2016NOMA}\\
\cline{2-5}
               &\multirow{3}{*}{Power Allocation}  &$\bullet$ Geometric Programming & Low complexity  & \cite{Boya2016NOMA} \\
\cline{3-5}
               &  &$\bullet$ Lagrangian Algorithms& Closed-form solutions   & \cite{Ali2017MIMONOMA} \\
 \cline{3-5}
               &  &$\bullet$ Bi-section search& Low complexity  & \cite{Yuanwei2016NOMA} \\
\hline

\end{tabular}}
\end{center}
\caption{Summary of representative optimization approaches for resource allocation in MIMO-NOMA}\label{tab2}
\end{table}

 \end{document}